\documentclass[sigconf]{acmart}

\AtBeginDocument{%
  }

\usepackage{multirow}
\usepackage{subcaption}

\copyrightyear{2026}
\acmYear{2026}
\setcopyright{cc}
\setcctype{by}
\acmConference[AST '26]{7th ACM/IEEE International Conference on Automation of Software Test (AST 2026)}{April 13--14, 2026}{Rio de Janeiro, Brazil}
\acmBooktitle{7th ACM/IEEE International Conference on Automation of Software Test (AST 2026) (AST '26), April 13--14, 2026, Rio de Janeiro, Brazil}
\acmPrice{}
\acmDOI{10.1145/3793654.3793752}
\acmISBN{979-8-4007-2476-3/2026/04}

\begin{document}

\title{Understanding Bug-Reproducing Tests: A First Empirical Study}

\author{Andre Hora}
\orcid{0000-0003-4900-1330}
\affiliation{%
  \institution{Department of Computer Science, UFMG}
  \city{Belo Horizonte}
  \country{Brazil}
}
\email{andrehora@dcc.ufmg.br}

\author{Gordon Fraser}
\orcid{0000-0002-4364-6595}
\affiliation{%
  \institution{University of Passau}
  \city{Passau}
  \country{Germany}
}
\email{gordon.fraser@uni-passau.de}

\begin{abstract}
Developers create bug-reproducing tests that support debugging by failing as long as the bug is present, and passing once the bug has been fixed.
These tests are usually integrated into existing test suites and executed regularly alongside all other tests to ensure that future regressions are caught.
Despite this co-existence with other types of tests, the properties of bug-reproducing tests are scarcely researched, and it remains unclear whether they differ fundamentally.
In this short paper, we provide an initial empirical study to understand bug-reproducing tests better.
We analyze 642 bug-reproducing tests of 15 real-world Python systems.
Overall, we find that bug-reproducing tests are not (statistically significantly) different from other tests regarding LOC, number of assertions, and complexity.
However, bug-reproducing tests contain slightly more try/except blocks and ``weak assertions'' (e.g.,~\texttt{assert\-Not\-Equal}).
Lastly, we detect that the majority (95\%) of the bug-reproducing tests reproduce a single bug, while 5\% reproduce multiple bugs.
We conclude by discussing implications and future research directions.
\end{abstract}

%
%




\maketitle

\section{Introduction}

When fixing a bug, ideally, developers should create a corresponding automated test that reproduces the bug, ensuring that this test fails when the buggy code is present and passes once the bug has been fixed~\cite{thomas2019pragmatic, winters2020software}.
This best practice ensures that future regressions are caught and is widely recommended: \emph{``The best way to start fixing a bug is to make it reproducible. After all, if you can't reproduce it, how will you know if it is ever fixed?''}~\cite{thomas2019pragmatic}.
This practice is also common in open-source projects~\cite{next, black}, for example, the contribution guidelines of the Black project state: \emph{``If you’re fixing a bug, add a test. Run it first to confirm it fails, then fix the bug, run it again to confirm it’s really fixed''}~\cite{black}.

\begin{figure}[t]
     \centering
     \begin{subfigure}[b]{0.48\textwidth}
         \centering
         \includegraphics[width=\textwidth]{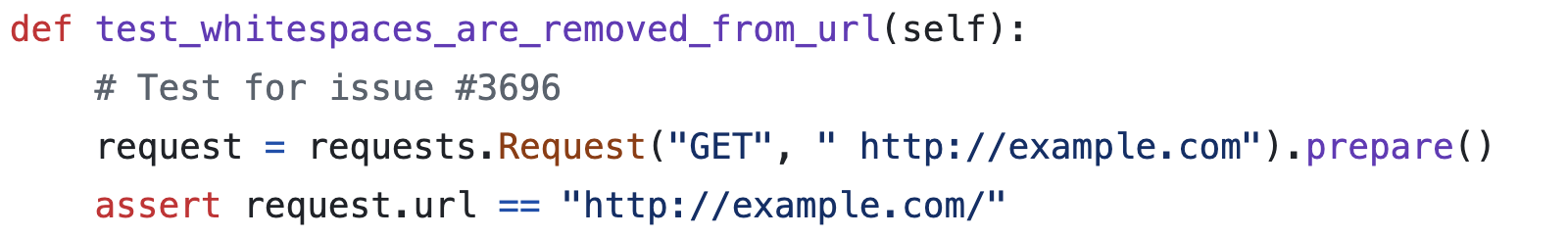}
         \caption{Test in Requests (bug \#3696).}
         \label{fig:ex1a}
     \end{subfigure}
     \par\medskip
     \begin{subfigure}[b]{0.48\textwidth}
         \centering
         \includegraphics[width=\textwidth]{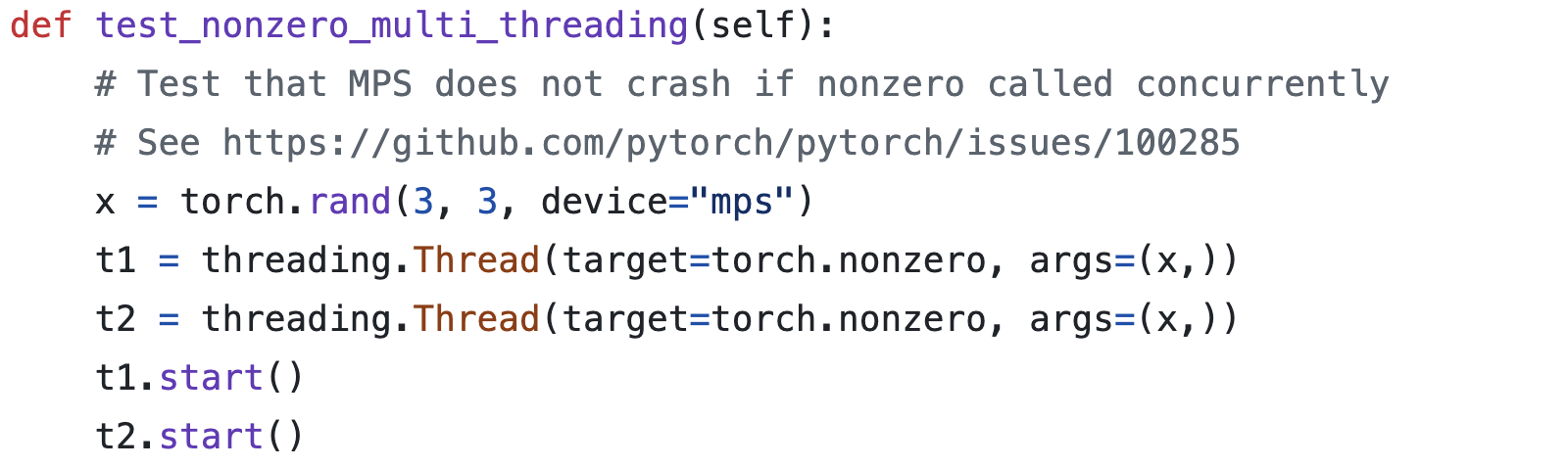}
         \caption{Test in PyTorch (bug \#100285).}
         \label{fig:ex1c}
     \end{subfigure}
    \caption{Examples of bug-reproducing tests.}
    \Description{Examples of bug-reproducing tests.}
    \label{fig:examples}
\end{figure}

Figure~\ref{fig:ex1a} shows a real example of a bug-reproducing test: it contains a bug ID that links it to a description of the underlying bug, which causes whitespaces to not be correctly removed from URLs (project ``requests'', bug \#3696).\footnote{Requests: \url{https://github.com/psf/requests/blob/79b74ef/tests/test_requests.py\#L187}; \url{https://github.com/psf/requests/issues/3696}} The test consists of a call to the API similar to an example given in the bug report. 
As another example, the test in Figure~\ref{fig:ex1c} additionally contains a short description, and then exercises concurrent usage on a certain GPU, though this time using an artificial scenario derived by the developer, rather than the actual user-reported call (PyTorch, bug \#100285).\footnote{PyTorch: 
\url{https://github.com/pytorch/pytorch/blob/d1f73fd/test/test_mps.py\#L10786}; \url{https://github.com/pytorch/pytorch/issues/100285}}
Note that this test has no assertions, meaning it will fail only if the production code raises an exception~\cite{hora2025exceptional, dalton2020exceptional, marcilio2021java}.
Both tests focus on testing individual bugs, but bug-reproducing tests may also test multiple bugs, as exemplified by the CPython test \texttt{test\_\-strftime},\footnote{CPython: \url{https://github.com/python/cpython/blob/d13ee0ae186f4704f3b6016dd52f7727b81f9194/Lib/test/datetimetester.py\#L1586}} which reproduces five bugs.
There are also cases where multiple tests are needed to reproduce a single bug properly.
For instance, four tests have been created in the Pandas project to reproduce bug \#3490.\footnote{Pandas: \url{https://github.com/pandas-dev/pandas/blob/a2fb11e/pandas/tests/plotting/test_datetimelike.py\#L1445-L1508}}


Besides their role in software development, bug-reproducing tests are also frequently used in research, for example, in databases of real bugs~\cite{just2014defects4j, widyasari2020bugsinpy, tomassi2019bugswarm}, program repair~\cite{hu2019re, monperrus2018automatic}, test reduction~\cite{regehr2012test, kreutzer2021test}, and negative/exceptional tests~\cite{hora2024test, hora2025exceptional, dalton2020exceptional, marcilio2021java}.

In practice, bug-reproducing tests are usually integrated into existing test suites and executed regularly alongside all other tests to ensure that future regressions are caught.
\textbf{Despite this co-existence with other types of tests, the properties of bug-reproducing tests are scarcely researched, and it remains unclear whether they differ fundamentally.}

In this paper, we provide an initial empirical study to understand bug-reproducing tests better, in order to provide a foundation for assessing whether such tests adhere to best testing practices and identifying potential areas for improvement.
We analyze 642 bug-reproducing tests of 15 real-world Python systems, addressing two research questions:

\smallskip
   
\noindent\textbf{RQ1: What are the code characteristics of bug-reproducing tests?}
Overall, we find that bug-reproducing tests are not (statistically significantly) different from other tests regarding LOC, number of assertions, and complexity. 
However, bug-reproducing tests have slightly more try/except blocks and ``weak assertions''.

\smallskip

\noindent\textbf{RQ2: How are bugs mapped to bug-reproducing tests?}
The majority (95\%) of the bug-reproducing tests reproduce a single bug, while a minority (5\%) reproduce multiple bugs. Moreover, sometimes, multiple tests are needed to reproduce a bug. We find that 20\% of the bug-reproducing tests reproduce shared bugs.

\smallskip
   
\noindent\textbf{Contributions.}
The contributions of this study are twofold.
First, we describe the first empirical study to understand bug-reproducing tests in the wild, in real-world systems.
Second, we discuss actionable implications and future research direction.

\section{Study Design}

\subsection{Case Studies}

We aim to study bug-reproducing tests of real-world and relevant systems.
For this purpose, we selected the top-15 most popular Python software systems hosted on GitHub according to the number of stars, a metric primarily adopted in the software mining literature as a proxy of popularity~\cite{hudson16, jss-2018-github-stars}.
We focus on Python because it is the most popular programming language nowadays, and it has a rich software ecosystem.
The 15 selected systems are Transformers (132K stars), TheFuck (85K), PyTorch (82K), Django (80K), FastAPI (76K), Flask (68K), Ansible (62K), CPython (62K), Scikit-Learn (60K), Requests (52K), Scrapy (52K), Rich (50K), Pandas (43K), Black (39K), and Sentry (39K).
In total, these systems have 121,447 test methods.
Our dataset is available at: \url{https://doi.org/10.5281/zenodo.17468629}.

\subsection{Detecting Bug-Reproducing Tests}


Bugs are typically reported and stored in Issue Tracking Systems, such as GitHub Issues.
These systems manage not only bugs but also other reported issues, such as new features, refactorings, and more.
Labels such as \emph{bug}, \emph{new feature}, and \emph{refactoring} can be used to categorize issues, but they are optional.
Ideally, a bug-reproducing test should be linked to its corresponding bug-labeled issue.
Links between code and issues can be found in various artifacts, such as pull requests (PRs), commits, and code comments.
Using such indicators, starting from a bug-labeled issue, we can find its linked artifacts (such as commits or PRs) that actually contain the bug-reproducing tests.
We can also perform the opposite operation: we can start from a commit or a PR, detect issue IDs in commit/PR messages (e.g.,~\emph{``fixed issue \#123''}), verify whether the issue is really a bug, and then identify bug-reproducing tests.
However, both solutions present some drawbacks:
(1) issues are not necessarily properly labeled;
(2) commit messages may not include issue IDs; and
(3) commits may include unrelated changes~\cite{dias2015untangling} (e.g.,~a bug fix and a new feature).

To avoid these possible limitations, this study focuses on detecting tests that the developers themselves label in the source code as bug-reproducing.
This is a conservative method where we prioritize precision over recall.
This solution is inspired by the rich literature on self-admitted technical debt, which has successively used a similar motivation to detect technical debt~\cite{potdar2014exploratory, sierra2019survey}.
For this purpose, we mine test methods that directly include the words \emph{``bug''} or \emph{``regression''} and refer to an issue ID in test comments or test names.
For example, test methods with comments like ``\emph{\# Regression \#123}'' or with test names like \texttt{test\_bug\_123}.

Following this method identifies 642 bug-reproducing tests in 10 out of the 15 analyzed systems, as detailed in Table~\ref{tab:bugs}.
We note that five projects have a higher number of bug-reproducing tests: CPython (264), Django (149), Scikit-Learn (81), PyTorch (64), and Pandas (55).
Other projects have lower numbers: Transformers~(16), Rich (5), Sentry (5), Black (2), and Scrapy (1).
Also, 23\% (150) of the bug-reproducing tests include the bug ID in their names, e.g.,~\texttt{test\_\-bug\_\-3061}.\footnote{CPython: \url{https://github.com/python/cpython/blob/d13ee0ae186f4704f3b6016dd52f7727b81f9194/Lib/test/test_time.py\#L657}} 




\begin{table}[t]
    \centering
    \small
    \caption{Bug-reproducing tests by project.}
    \begin{tabular}{lrrr}
        \toprule
        \multirow{2}{*}{\textbf{Project}} & \multirow{2}{*}{\textbf{Total}} & \multicolumn{2}{c}{\textbf{Bug ID in...}} \\ \cline{3-4}
        & & \textbf{Test Name} & \textbf{Test Comment} \\
        \midrule
            CPython	&	264	&	99	&	165	\\
            Django	&	149	&	28	&	121	\\
            Scikit-Learn	&	81	&	0	&	81	\\
            PyTorch	&	64	&	1	&	63	\\
            Pandas	&	55	&	21	&	34	\\
            Transformers	&	16	&	0	&	16	\\
            Rich	&	5	&	0	&	5	\\
            Sentry	&	5	&	1	&	4	\\
            Black	&	2	&	0	&	2	\\
            Scrapy	&	1	&	0	&	1	\\ \midrule
            Total	&	642	&	150 (23\%)	&	492 (77\%)	\\
        \bottomrule
    \end{tabular}
    \label{tab:bugs}
\end{table}

\subsection{Research Questions}



\subsubsection{RQ1: What are the code characteristics of bug-reproducing tests?}
We compute four metrics from the bug-reproducing tests: LOC (lines of code), number of assertions, complexity, and number of try/except blocks.
Complexity is measured by the count of control flow structures in the test code, such as \texttt{if}, \texttt{for}, \texttt{while}, and \texttt{try}.
For comparison, we also compute the same metrics for all 121K tests of the 15 selected systems.
We apply the Mann-Whitney U-test and Cohen effect size to verify whether the metrics of both groups of tests differ.
We further explore differences in assertions by extracting the most used assertion commands in each group of tests.
We analyze the usage of ``weak assertions'', that is, assertions that are potentially less effective~\cite{MutationAnalysis}, such as \texttt{assert\-Not\-Equal} and \texttt{assert\-Contains}.
\textbf{Rationale:}
We aim to understand better whether bug-reproducing tests differ from ordinary tests regarding the analyzed metrics and whether they tend to use the same or distinct assertion commands.
Moreover, using ``weak assertions'' may indicate that the tested code is somehow harder to test.

\subsubsection{RQ2: How are bugs mapped to bug-reproducing tests?}

\begin{figure}[t]
    \centering
    \includegraphics[width=0.48\textwidth]{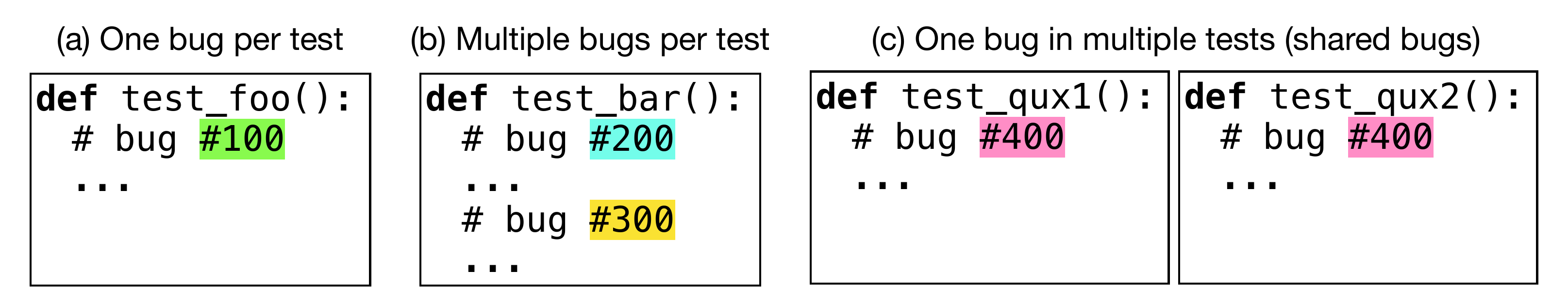}
    \caption{Mapping between bugs and bug-reproducing tests.}
    \Description{Mapping between bugs and bug-reproducing tests.}
    \label{fig:mapping}
\end{figure}

As detailed in Figure~\ref{fig:mapping}, we focus on three possible scenarios: (a) one bug per test, (b) multiple bugs per test, and (c) one bug in multiple tests (shared bugs, for short).
\textbf{Rationale:}
Our goal is to reveal how developers map bugs to tests.
Ideally, a test is focused and reproduces a single bug, as represented by scenario (a).
In this case, a failing test will clearly indicate the problem~\cite{winters2020software}.
In contrast, cases like (b) are less desirable because the test reproduces multiple bugs (like a test verifying multiple functionalities).
In this case, the test may be harder to understand, and a failing test will not clearly address the problem.
Lastly, cases like (c) may indicate that a bug is too complex to be revealed or checked by a single test, like a larger functionality that is broken into smaller, testable ones. 

\section{Results}

\subsection{RQ1: Code Characteristics}

Table~\ref{tab:metrics} details the code metrics extracted from the bug-reproducing tests, as well as for all 121K test methods of the 15 analyzed projects (column \emph{All Tests}).
We find that the differences are statistically significant (\emph{p-value $<$ 0.01}) for LOC, but the difference has a \emph{negligible} effect size, and the difference is significant for try/excepts with a \emph{small} effect size.
This means that there are no practical differences between bug-reproducing and all tests in LOC, assertions, and complexity, while there are significant/small differences for try/excepts.
\begin{table}[t]
    \centering
    \scriptsize
    \caption{Code characteristics of bug-reproducing tests. Q1, Q2, Q3: first, second, and third quartiles. ES: effect-size}
    \begin{tabular}{lr rr rl}
        \toprule
        \multirow{2}{*}{\textbf{Metric}} & \multirow{2}{*}{\textbf{Quartile}} & \textbf{All} & \textbf{Bug-Reproducing} & \multirow{2}{*}{\textbf{p-value}} & \multirow{2}{*}{\textbf{ES}} \\ 
        & & \textbf{Tests} & \textbf{Tests} & & \\ \midrule
        \multirow{4}{*}{LOC}                 & Q1 & 5 & 6 & \multirow{4}{*}{$<$0.01} &    \\
        & Q2 & 10 & 11 & & $+$0.064  \\
        & Q3 & 20 & 19 & & negligible \\
        & Mean & 16 & 17.8 & \\ \midrule
        \multirow{4}{*}{Assertions}           & Q1 & 1 & 1 & \multirow{4}{*}{0.32} &   \\
        & Q2 & 2 & 2 & & $+$0.103 \\
        & Q3 & 4 & 3 & & negligible \\
        & Mean & 3 & 3.5 & \\ \midrule
        \multirow{4}{*}{Complexity}         & Q1 & 0 & 0 & \multirow{4}{*}{0.16} &  \\
        & Q2 & 0 & 0 & & $-$0.003 \\
        & Q3 & 0 & 0 & & negligible \\
        & Mean & 0.39 & 0.38 & \\ \midrule
        \multirow{4}{*}{Try/except}         & Q1 & 0 & 0 & \multirow{4}{*}{$<$0.01} &  \\
        & Q2 & 0 & 0 & & $+$0.265 \\
        & Q3 & 0 & 0 & & small \\
        & Mean & 0.029 & 0.088 & \\
        \bottomrule
    \end{tabular}
    \label{tab:metrics}
\end{table}

\textbf{Observation 1}:
Overall, bug-reproducing tests are not (statistically significantly) different from other tests regarding the number of lines of code, number of assertions, and complexity.
However, bug-reproducing tests tend to have slightly more try/except blocks.


Despite the similarities, considering the overall comparison, we also find an important difference between the groups.
First, 6\% of the bug-reproducing tests have at least one try/except block, while this value for all tests is only 2\%.
Second, despite having an equivalent number of assertions, they are not necessarily the same.
To further investigate this potential difference, Table~\ref{tab:assertions} details the top-25 most commonly used assertion commands in both all tests and bug-reproducing tests.
We paid special attention to the usage of ``weak assertions'', that is, assertions that are potentially less effective~\cite{MutationAnalysis}, such as \texttt{assert\-Not\-Equal}, \texttt{assert\-Almost\-Equal}, and \texttt{assert\-Contains}.
We find 8/25 weak assertions in all tests (from which 2 are exclusives), while 11/25 in bug-reproducing tests (from which 4 are exclusives).



\begin{table}[t]
    \centering
    \scriptsize
    \caption{Top-25 most commonly used assertions (legend: \textcolor{red}{weak assertions} and \textcolor{red}{\colorbox{yellow}{exclusive weak assertions}}).}
    \begin{tabular}{p{8cm}}
        \toprule
        
        \multicolumn{1}{c}{\textbf{All Tests (8 weak assertions, 2 exclusives)}} \\ \midrule
        
        assertEqual, assert, assertTrue, assert\_frame\_equal, assertFalse, \textcolor{red}{assertIn}, assertIs, assert\_series\_equal, \textcolor{red}{assertIsInstance}, assert\_index\_equal, assertIsNone, assertListEqual, assert\_numpy\_array\_equal, \textcolor{red}{assertNotEqual}, assertSequenceEqual, \textcolor{red}{assertNotIn}, \textcolor{red}{\colorbox{yellow}{assertIsNotNone}}, assertExpectedInline, \textcolor{red}{assertContains}, assert\_equal, \textcolor{red}{assert\_called\_with}, assert\_produces\_warning, \textcolor{red}{\colorbox{yellow}{assertAlmostEqual}}, assertNumQueries, assertQuerySetEqual \\ \toprule
        
        \multicolumn{1}{c}{\textbf{Bug-Reproducing Tests (11 weak assertions, 4 exclusives)}} \\ \midrule
        
        assertEqual, assert, assertTrue, assertIs, assertQuerySetEqual, \textcolor{red}{assertIsInstance}, assertSequenceEqual, assertFalse, assertNumQueries, assert\_index\_equal, assertIsNone, \textcolor{red}{assertIn}, assert\_series\_equal, assert\_frame\_equal, \textcolor{red}{assertContains}, \textcolor{red}{\colorbox{yellow}{assertNotHasAttr}}, \textcolor{red}{assertNotIn}, assertRedirects, assert\_numpy\_array\_equal, \textcolor{red}{assertNotEqual}, \textcolor{red}{\colorbox{yellow}{assertNotContains}}, \textcolor{red}{assertIsNot}, \textcolor{red}{assert\_called\_with}, \textcolor{red}{\colorbox{yellow}{assertCountEqual}}, \textcolor{red}{\colorbox{yellow}{assertLess}} \\
        
        \bottomrule
    \end{tabular}
    \label{tab:assertions}
\end{table}

\textbf{Observation 2}:
Bug-reproducing tests contain slightly more ``weak assertions'' (e.g.,~\texttt{assert\-Not\-Equal} and \texttt{assert\-Contains}) than other tests.
6\% of the bug-reproducing tests have try/except blocks.

\subsection{RQ2: Bug to Test Mapping}

Table~\ref{tab:mapping} shows that 95\% of the bug-reproducing tests handle a single bug, while 5\% handle multiple bugs.
For example, the two bug-reproducing tests presented in Figure~\ref{fig:examples} focus on a single bug.
In contrast, the Django test \texttt{test\_\-more\_\-more\_\-more}\footnote{Django: \url{https://github.com/django/django/blob/790f0f8/tests/aggregation_regress/tests.py\#L1026}} reproduces six bugs (\#10199, \#10248, \#10290, \#10425, \#10666, and \#10766).

\begin{table}[t]
    \centering
    \footnotesize
    \caption{Mapping between bugs and bug-reproducing tests.}
    \begin{tabular}{llrr}
        \toprule
        && \textbf{\#} & \textbf{\%} \\ \midrule
        \multirow{2}{*}{One/Multiple}& Tests reproducing one bug & 613 & 95\% \\
        & Tests reproducing multiple bugs       & 29 & 5\% \\ \midrule
        \multirow{2}{*}{Exclusive/Shared}       & Tests reproducing exclusive bugs    & 508 & 80\%  \\
        &  Tests reproducing shared bugs        & 134 & 20\%\\
        \bottomrule
    \end{tabular}
    \label{tab:mapping}
\end{table}

\textbf{Observation 3}:
The majority (95\%) of the bug-reproducing tests reproduce a single bug; a minority (5\%) reproduce multiple bugs.

We also explore whether the bugs are exclusive or shared among the bug-reproducing tests (Table~\ref{tab:mapping}).
We find that 80\% of the tests reproduce exclusive bugs, that is, bugs that are tested only by a single test.
In contrast, 20\% of tests reproduce shared bugs, that is, bugs that are tested by multiple tests.
For example, the bug \#11371 in Django is tested by \texttt{test\_post} and \texttt{test\_put}.\footnote{Django: \url{https://github.com/django/django/blob/790f0f8/tests/test_client_regress/tests.py\#L1083-L1101}}

\textbf{Observation 4}:
Sometimes, multiple tests are needed to reproduce a single bug.
We find that 20\% of the bug-reproducing tests reproduce shared bugs.

\section{Discussion}

\noindent\textbf{Improve tests with weak assertions and try/except blocks.}
Overall, we detected that bug-reproducing tests contain slightly more try/except blocks and ``weak assertions'' than other tests.
Verifying exception-raising via \texttt{assertRaises} commands is a better choice than try/except blocks to properly assert that an exception is raised.
Moreover, using strong assertions is a better choice to verify the program's output more effectively.
Several reasons may lead to the use of such weak assertions and try/except blocks, such as developer inattention or vague user-written bug reports.
Another possible explanation is that such tests suffer from poor observability~\cite{aniche_blog}.
In this context, tests without assertions may happen due to this lack of observability: when it is hard to observe the program's output, developers cannot write assertions~\cite{aniche_blog}.
We hypothesize that some bug-reproducing tests face similar limitations, lacking straightforward methods for observing outputs, which leads to the use of weak assertions or try/except blocks as a workaround.
\emph{Future research could explore such tests, proposing solutions to enhance observability and recommending the usage of strong assertions.}

\smallskip

\noindent\textbf{Split multi-bug test into multiple single-bug tests.}
Ideally, a test should reproduce a single bug so that a failing test indicates the problem~\cite{winters2020software}.
A test that reproduces multiple bugs is harder to understand, and its failure will not clearly address the problem.
We found that the majority of tests target a single bug (95\%), but some tests target multiple bugs (5\%).
\emph{This can be an opportunity to support developers (e.g.,~via contribution study~\cite{brandt2024shaken, hora2024pathspotter, danglot2019automatic}) by investigating the possibility of turning one multi-bug test into multiple single-bug tests.}

\smallskip

\noindent\textbf{Avoid naively replicating bug-reproducing scenarios.}
While performing our study, we noticed multiple bug-reproducing tests that originated almost entirely from bug-reproducing scenarios reported by users in the issue tracker.
For instance, the PyTorch bug \#116095\footnote{\url{https://github.com/pytorch/pytorch/issues/116095}} presents a bug scenario that is replicated in test \texttt{test\_\-cross\_\-entropy\_\-loss}.\footnote{\url{https://github.com/pytorch/pytorch/blob/d1f73fd844/test/test_mps.py\#L4860}}
Similarly, the CPython bug \#620179\footnote{\url{https://bugs.python.org/issue620179}} includes a bug scenario that ends up in \texttt{test\_ipow}.\footnote{\url{https://github.com/python/cpython/blob/4767a6e31c055/Lib/test/test_descr.py\#L3973}}
When creating test cases, ideally, developers should strive to minimize tests for bug reproduction, a technique known as \emph{test reduction}~\cite{regehr2012test, kreutzer2021test, gcc_test, llvm_test, webkit_test}.
A reduced test can help identify the central problem, allowing developers to spend more time determining the solution~\cite{webkit_test}.
In this context, one potential problem happens when developers copy and paste bug-reproducing scenarios directly into tests.
For example, in the two examples provided, neither test included assertions; instead, they simply replicated the bug scenario as reported.
This approach misses the opportunity to validate expected outputs and confirm that the bug has been fixed.
\emph{Future research could investigate methods to identify this practice and warn developers about potentially less effective bug-reproducing tests.}

\section{Limitations}

We analyzed a large set of bug-reproducing tests, but they do not represent all possible bug-reproducing tests in the selected systems.
Our heuristic detected bug-reproducing tests that are explicitly admitted by developers in the source code.
Therefore, further studies could increase recall and expand the set of bug-reproducing tests.

\section{Related Work}

Bug reproducing tests are related to multiple aspects of testing research.
For example, \emph{databases of real bugs} (e.g.,~Defects4J~\cite{just2014defects4j}, BugsInPy~\cite{widyasari2020bugsinpy}, and BugSwarm~\cite{tomassi2019bugswarm}) are typically accompanied by bug-reproducing tests.
\emph{Program repair techniques} may rely on failing tests to create patches that make the test suite pass~\cite {hu2019re, monperrus2018automatic}.
Creating a bug-reproducing test is also part of the process of \emph{test reduction}~\cite{regehr2012test, kreutzer2021test}, a technique that aims to minimize test cases for bug reproduction.
The importance of test reduction is highlighted in projects such GCC~\cite{gcc_test}, LLVM~\cite{llvm_test}, and WebKit~\cite{webkit_test}.
In those studies, the main focus is on the buggy/fixed code, while bug-reproducing tests are present to support that the fixed code works properly.
Lastly, bug reproducing tests are also related to a set of studies in the context of testing negative and exceptional tests~\cite{hora2024test, hora2025exceptional, dalton2020exceptional, marcilio2021java} and may support a better understanding of them.

\section{Conclusions and Future Work}

We provided an initial empirical study to understand bug-reproducing tests better.
Overall, we found that bug-reproducing tests are not different from other tests regarding LOC, number of assertions, and complexity, but they contain slightly more try/except blocks and ``weak assertions''.
We also found that the majority (95\%) of the bug-reproducing tests reproduce a single bug, while 5\% reproduce multiple bugs.
We concluded by discussing multiple possibilities to improve bug-reproducing tests.

\smallskip

\noindent\textbf{Future Work:}
First, this study assesses syntactic differences between bug-reproducing tests and others.
Future work could look at semantic properties, e.g.,~do they differ in terms of coverage achieved per test or mutants killed?
Second, understanding the differences between bug reproducing and ``regular'' tests is also important to automatically generate such tests, e.g.,~by developing adequate prompts that reflect these properties when using LLMs to generate tests~\cite{fan2023large, hou2023large, schafer2023empirical, alshahwan2024automated, ouedraogo2024test}.
Finally, as testing is generally well adopted in the industry, some companies invest huge amounts of resources into test execution infrastructure.
In this context, one important question is whether bug-reproducing tests and other tests need to be executed at the same frequency, which may lead to a potential reduction of cost.


\begin{acks}
This research was supported by CNPq (process 403304/2025-3), CAPES, and FAPEMIG.
This work was partially supported by INES.IA (National Institute of Science and Technology for Software Engineering Based on and for Artificial Intelligence), www.ines.org.br, CNPq grant 408817/2024-0. 
\end{acks}

\bibliographystyle{ACM-Reference-Format}
\bibliography{main}

\end{document}